\documentclass[aps,pra,twocolumn,groupedaddress,amsmath,amsfonts,amssymb]{revtex4-1}
\usepackage[english]{babel}
\usepackage{ucs}
\usepackage[utf8x]{inputenc}
\usepackage{graphicx}
\usepackage{bm}
\usepackage{bbm}
\usepackage{empheq}

\newcommand{\ue}{\mathrm{e}}

\begin{document}
\title{Role of hyperfine interaction for cavity-mediated coupling between spin qubits}

\author{Julia Hildmann and Guido Burkard}

\affiliation{Department of Physics, University of Konstanz, D-78457 Konstanz, Germany}

\begin{abstract}
We consider two qubits interacting by means of an optical cavity, where each qubit is
represented by a single electron spin confined to a quantum dot.
It is known that electron spins in III-V semiconductor quantum dots are affected by the decoherence due to 
the hyperfine interaction with nuclear spins. Here we show that the interaction between two qubits is 
influenced by the Overhauser field as well.
Starting from an unpolarizied nuclear ensemble, we investigate the 
dependance of the fidelities for two-qubit gates
on the Overhauser field. We include the hyperfine interaction perturbatively to second order in our
analytical results, and to arbitrary precision numerically.

\end{abstract}

\maketitle

\section{Introduction}
\begin{figure}
\includegraphics[width=0.45\textwidth]{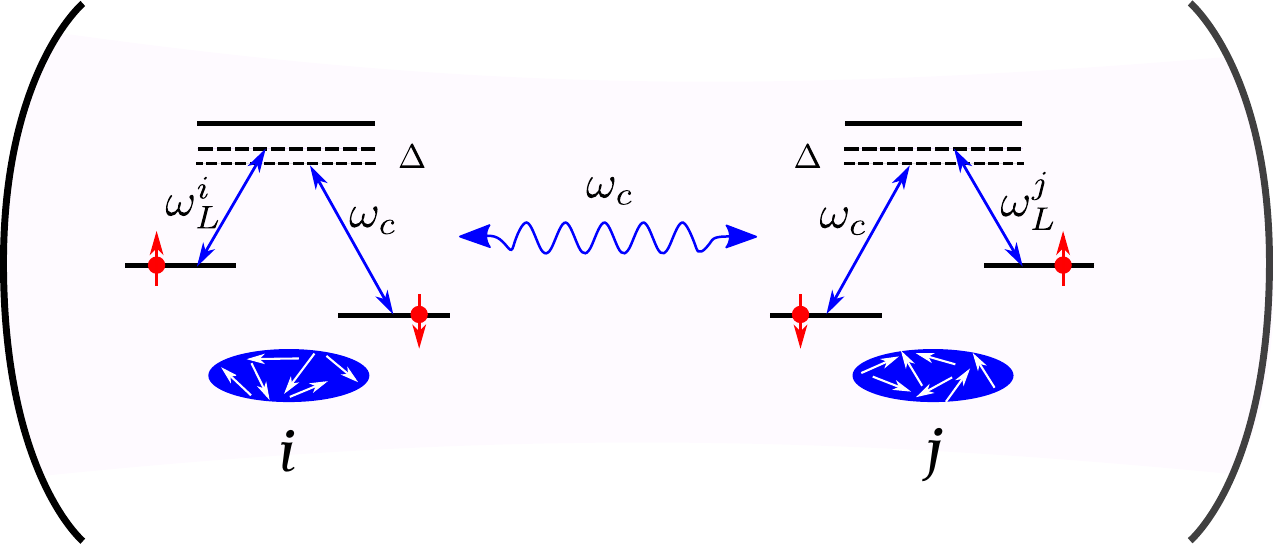}
\caption{Cavity mode mediated spin-spin interaction. If the difference of the
detunings $\Delta$ for two quantum dots coincides, then they can exchange virtual cavity photons of
energy $\omega_c$. Here, $\omega_c$ and  $\omega_L^{i}$ are the frequencies of the cavity mode
and the laser acting on quantum dot $i$, and $\Delta$ denotes the difference between the detunings,
i.e. the detuning from the two-photon resonance on a given quantum dot.}
\label{schematic}
\end{figure}

Substantial progress has been made towards the implementation of 
coherent interfaces between an electron 
spin in a quantum dot and photons \cite{mikkelsen_science2008, berezovsky_science2008, yilmaz_prl2010, Liu_adv2010}. 
The coupling of single spins to photons is a promising mechanism for implementing
quantum information processing schemes based on quantum dots, where 
a qubit (quantum bit of information) is represented by electron spin degrees of freedom 
\cite{loss_divicenzo_proposaL_QD}.
The main source of electron spin decoherence in semiconductor quantum dots is considered
to be the hyperfine interaction to the nuclear spins of the lattice atoms 
\cite{khaetskii_prl2002,coish_statsolidi2009}.
The dephasing time due to the interaction with an unprepared ensemble of nuclear spins is about 10 ns 
\cite{petta_science2005, coish_statsolidi2009}, which is much shorter than the single qubit operation 
times by means of electronic control ($\approx$ 100 ns \cite{koppens_nature2006}).   
There are some ways to deal with this problem: to prolong the decoherence time 
using advanced spin-echo techniques \cite{bluhm_natphys2010} or dynamical nuclear polarization
\cite{foletti_natphys2009}, to use nuclear spin ``poor" materials such as 
graphene \cite{trauzettel_natphys2007} or  nitrogen-vacancy centers in diamond \cite{hanson_nature2008}
or to shorten the qubit manipulation time \cite{hugo_2010}. The optical driving of 
electron spins has this last advantage and offers mechanisms for single qubit operation time around
30 ps \cite{berezovsky_science2008}. Besides single qubit manipulations, two qubit operations are
required for implementing quantum algorithms. 
One of the schemes for optical exchange interaction between 
two spins is based on virtual Raman transitions between  valence band heavy holes  and 
conduction band electrons \cite{imamoglu_prl99}.
Thereby, a set of quantum dots is coupled to a high finesse cavity.
Each quantum dot is doped with a single electron and
can be controlled by linearly polarized laser light. 
The laser polarization is perpendicular 
to the growth direction of the quantum dots and 
perpendicular to the cavity field polarization as well, 
so that laser light is acting on the quantum dots without
driving the cavity. A magnetic field is applied in the Voigt configuration, i.e.,
perpendicular to the growth direction and parallel to the cavity field polarization,
so that single photons emited by quantum dots escape 
into the cavity and can be detected afterwards. 
The cavity frequency is chosen such that it is
slightly detuned from the transition from 
the electron spin down state to the trion state, 
while the laser frequency is  detuned from the 
transition from the spin up state to the trion state.
If these detunings are the same then the $\Lambda$ system is
in a two-photon resonance that allows for manipulating a single electron spin 
(Figs.~\ref{schematic} and \ref{scheme}).
If the detunings in a quantum dot are different, but the difference of the detunings
(the detuning from the two-photon resonance) match
for a pair of quantum dots, then the electron spins in these quantum dots can 
resonantly exchange energy through virtual cavity photons (Fig. \ref{schematic}).
In this way, virtual photons can be used
for performing coherent interaction between arbitrary qubits. 
Two-qubit operations
can potentially be carried out on a subnanosecond timescale \cite{imamoglu_prl99}, 
therefore the decoherence can be expected to originate mainly from
nuclear spins. 
In this paper, we calculate the fidelities of such two-qubit operations
in the presence of the hyperfine interaction.
We will study in particular partial SWAP operations and the controlled-not (CNOT) gate. 
For this purpose, we first calculate the fidelity of the quantum gates of interest
analytically to second order of the Overhauser field and subsequently we perform
an exact  numerical calculation for comparison.

\section{Model}
We start from the microscopic model describing a quantum dot embedded in a cavity and
irradiated by a laser field,
\begin{eqnarray}
H &=& \sum_{\sigma =\uparrow,\downarrow,\pm 3/2}\omega_\sigma e_\sigma^\dagger e_\sigma
+\omega_c a^\dagger_c a_c + \omega_L a^\dagger_L a_L  + H_{\rm hf}\nonumber\\
& & + g_c a_c^\dagger e_{v}^\dagger e_{\uparrow} - ig_L a_L^\dagger
e_{v}^\dagger e_{\downarrow} + h.c.,
\label{eh-Hamiltonian}
\end{eqnarray}
where $\omega_{\sigma}$ and $e_\sigma$ are the energy and annihilation operators for 
the electronic states in the conduction band with spin $\sigma = \pm 1/2 = \uparrow, \downarrow$ 
and in the valence band with total angular momentum $\sigma = \pm 3/2$ (we assume that the
heavy hole subband is sufficiently split from the light hole band to allow for pure heavy hole
excitation). 
The quantization axis for the electrons is set by the external magnetic field along the
$x$ direction, while the quantization axis in the valence band is given by the structure and
assumed to be in $z$ direction.  Furthermore, for a linearly
polarized cavity mode ($\omega_c$, $a_c$)  along the $x$ direction and
linearly polarized laser mode ($\omega_L$,$a_L$) in the $y$ direction, we obtain
a radiative coupling to the linear combination $e_v = (e_{-3/2}-e_{3/2})/\sqrt{2}$ in the 
valence band.  The coupling strengths to the cavity and laser modes
are denoted with $g_c$ and $g_L$, respectively.
The hyperfine interaction of a conduction band electron in the quantum
dot with the surrounding nuclear spins can be written as
\begin{equation}
H_{\rm hf} = {\bf S}\cdot \sum_{k=1}^N A_k {\bf I}^k,
\end{equation}
where ${\bf I}^k$ is the operator belonging to the $k$-th nuclear
spin in contact with the electron and $A_k$ denotes the corresponding 
hyperfine coupling constant.  
The average hyperfine coupling constant for GaAs is 90 $\mu$eV
\cite{cerletti_nano2005}. 
The electron spin operator is given by
${\bf S} = \tfrac12\sum_{\sigma,\sigma'=\uparrow,\downarrow} e_\sigma^\dagger
\boldsymbol{\sigma}_{\sigma\sigma'}e_{\sigma'}$, where
$\boldsymbol{\sigma}=(\sigma_x,\sigma_y,\sigma_z)$ 
is the vector consisting of Pauli matrices.
Here, we neglect the dipolar hyperfine coupling of the valence band
states since it is typically smaller.
In the presence of a magnetic field along the $x$ direction exceeding
the nuclear field of typically $\sim 10\,{\rm mT}$, we can neglect the
flip-flop terms and write 
$H_{\rm hf}\approx  S_x \sum_{k=1}^N A_k I_x^k \equiv S_x h$,
where $h$ denotes the Overhauser (nuclear) field operator in $x$
direction. We do not take the nuclear Zeeman terms into account, because 
they are considerably smaller than the thermal energy and the electron Zeeman 
energy.
We can then combine the hyperfine Hamiltonian with the first term in
Eq.~(\ref{eh-Hamiltonian}) by using 
$\omega_\sigma = (g_e \mu_B B + h ) \sigma /2$
for $\sigma=\uparrow,\downarrow=\pm 1$,
where $B$ is the magnetic field applied along $x$
and $g_e$ is the effective electron g-factor.
We can now replace the operator $h$ with one of its eigenvalues 
and perform an average over $h$ later.  This allows us to
follow the steps performed in Ref.~\cite{imamoglu_prl99}
before taking the average over nuclear configurations.
In GaAs quantum dots the number $N$ of nuclear spins is
large, typically between $10^5$ and $10^6$, and therefore 
the Overhauser field follows a Gaussian distribution
around mean value 0 and with variance $\sigma\simeq A/\sqrt{N}$ \cite{cerletti_nano2005}.

\begin{figure}
\includegraphics[width=0.35\textwidth]{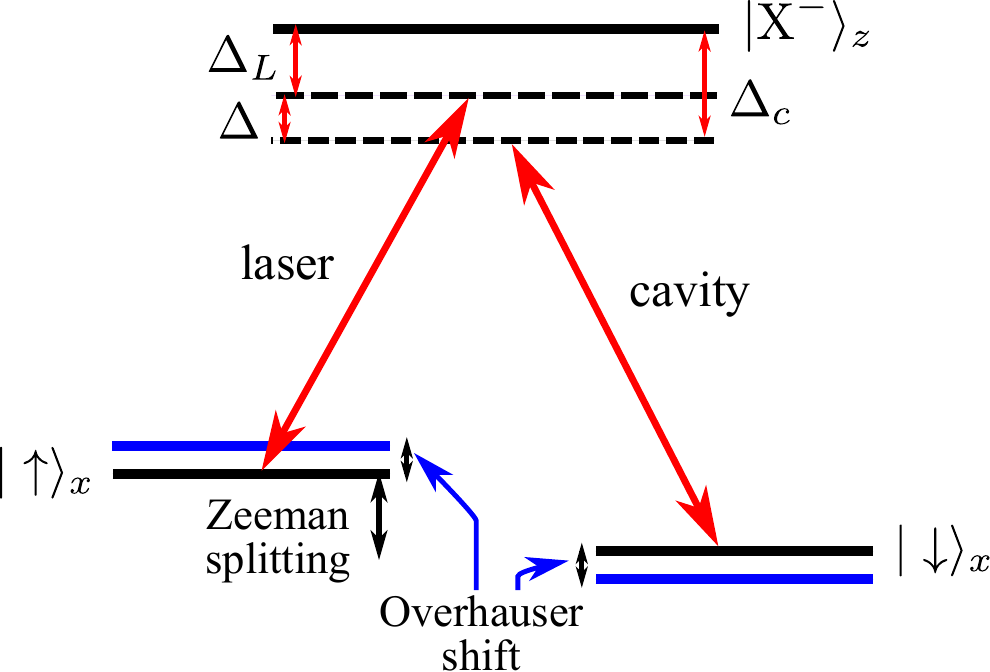}
\caption{Energy level scheme for a quantum dot filled with a single electron 
and coupled to a cavity mode. 
The Zeeman-split single-electron states can be excited to a trion (negatively charged exciton) state 
$|X^{-}\rangle$ 
by coupling to the cavity or laser field. 
Both cavity and laser field frequencies are detuned by $\Delta_c$ and $\Delta_L$ 
from resonance, and the combined system 
is detuned from its two-photon resonance by $\Delta = \Delta_c-\Delta_L$.
The Overhauser shift caused by the hyperfine coupling to the nuclear spins leads to 
a fluctuating detuning, thus reducing the fidelity of the optically generated quantum gates.}
\label{scheme}
\end{figure}

In order to eliminate the valence band states, we perform a
Schrieffer-Wolff transformation
$H_{\rm eff} = e^{-S}He^S$ with the anti-Hermitian operator
\begin{equation}
S = \frac{g_c}{\omega_\uparrow-\omega_v-\omega_c} a^\dagger_c e^\dagger_ve_\uparrow
- i\frac{g_L}{\omega_\downarrow-\omega_v-\omega_L} a^\dagger_L
e^\dagger_v e_\downarrow
- h.c.,
\end{equation}
where we use $\omega_{3/2}\approx\omega_{-3/2}\equiv \omega_v$.  
By setting the valence band occupation number  to $e^\dagger_v e_v\approx 1$ and
using that the  laser field is in a coherent state with $g_L a_L\approx \Omega_L
\exp(-i\omega_L t)$, we obtain  to second order in the cavity and
laser couplings $g_c$  and $g_L$,
\begin{eqnarray}
H_{\rm eff} = \omega_c a_c^\dagger a_c   &+&  \sum_{i=1,2} \Bigg[
(g_e\mu_B B + h_i)\sigma^i_{\uparrow\uparrow}  \label{Heff}\\
& &   +i g_{\rm eff}^i (a^\dagger_c  \sigma^i_{\downarrow\uparrow}e^{-i\omega_{L}^i t} -  h.c.)\nonumber\\
& &
 -\frac{g_c^2}{\Delta_c^i(h_i)}\sigma_{\downarrow\downarrow}^i
a_c^\dagger a_c 
-\frac{(\Omega_L^i)^2}{\Delta_L^i(h_i)}\sigma_{\uparrow\uparrow}^i\Bigg],\nonumber
\end{eqnarray}
where the index $i$ refers to the respective quantities in
dot  $i=1,2$, and $\sigma_{i\,j}=| i\rangle\langle
j|$,  $i, j=\uparrow,\downarrow$.
The first term inside the sum in Eq.~(\ref{Heff}) is the Zeeman
splitting due to the external and nuclear fields, the second term describes
the effective coupling of the cavity mode to the quantum dot electron
spins with strength
\begin{equation}
g_{\rm eff}^i(t)=\frac{g_c\Omega^i_L(t)}{2}\left(\frac{1}{\Delta^i_c(h_i)}+\frac{1}{\Delta^i_L(h_i)}\right),
\label{geff}
\end{equation}
where
\begin{eqnarray}
\Delta_c^i(h_i) &=& \frac{g_e\mu_B B + h_i}{2}-\omega_v-\omega_c   =
\Delta_c^i +\frac{h_i}{2},\label{Deltac}\\
\Delta_L^i (h_i)&=& -\frac{g_e\mu_B B + h_i}{2}-\omega_v-\omega_L^i   =
\Delta_L^i -\frac{h_i}{2} ,\label{DeltaL}
\end{eqnarray}
are the detunings of the cavity and laser fields (s. Fig. \ref{scheme}),
and the last two terms in Eq. (\ref{Heff}) can be interpreted as the Lamb and Stark shifts
of the cavity and quantum dot levels, respectively.

A second Schrieffer-Wolff transform can be used to also eliminate the 
cavity mode, which leads to 
the effective photon-mediated interaction between two spins $i$ and $j$
in the interaction picture with \\
$H_0=\sum_i (g_e\mu_B B+h_i)\sigma^i_{\uparrow\uparrow}$ ,
\begin{equation}
H_{\rm int}^{ij} = \frac{\tilde{g}_{ij}(t)}{2}(\sigma^i_{\uparrow\downarrow} \sigma^j_{\downarrow\uparrow}{\rm e}^{i(h_i-h_j)t} 
+ \sigma^j_{\uparrow\downarrow} \sigma^i_{\downarrow\uparrow}{\rm e}^{-i(h_i-h_j)t}),
\label{Hint}
\end{equation}
where 
\begin{equation}
\tilde{g}_{ij}(t)=\frac{g^i_{\rm eff}(t)g^j_{\rm eff}(t)}{2}\left(\frac{1}{\Delta_i(h_i)}+\frac{1}{\Delta_j(h_j)}\right), 
\label{gij}
\end{equation}
with
\begin{equation}
\Delta_i (h_i)= \Delta^i_c (h_i)-\Delta^i_L (h_i) = \Delta_i + h_i,
\label{Delta}
\end{equation}
represents the coupling strength between 
two electron spins. 
The time-independent interaction Hamiltonian Eq.~(\ref{Hint}) strictly
applies to the two-photon resonance $\Delta_i = \Delta_j \equiv \Delta$ in 
the absence of nuclear spins. By going over into 
the rotating frame with nuclear spins by 
$R=\exp{(i\,t(h_i\sigma^i_{\uparrow\uparrow}+h_j\sigma^j_{\uparrow\uparrow}))}$ 
we obtain a time-independent effective interaction Hamiltonian with Heisenberg transverse coupling type between two electron spins:
\begin{equation}
\tilde{H}^{\rm int}_{ij}=R^{\dagger}H_{\rm int}^{ij} R= \frac{\tilde{g}_{ij}(t)}{2}(\sigma^i_y\sigma^j_y+\sigma^i_z\sigma^j_z).
\end{equation}

\begin{figure}
\includegraphics[width=0.35\textwidth]{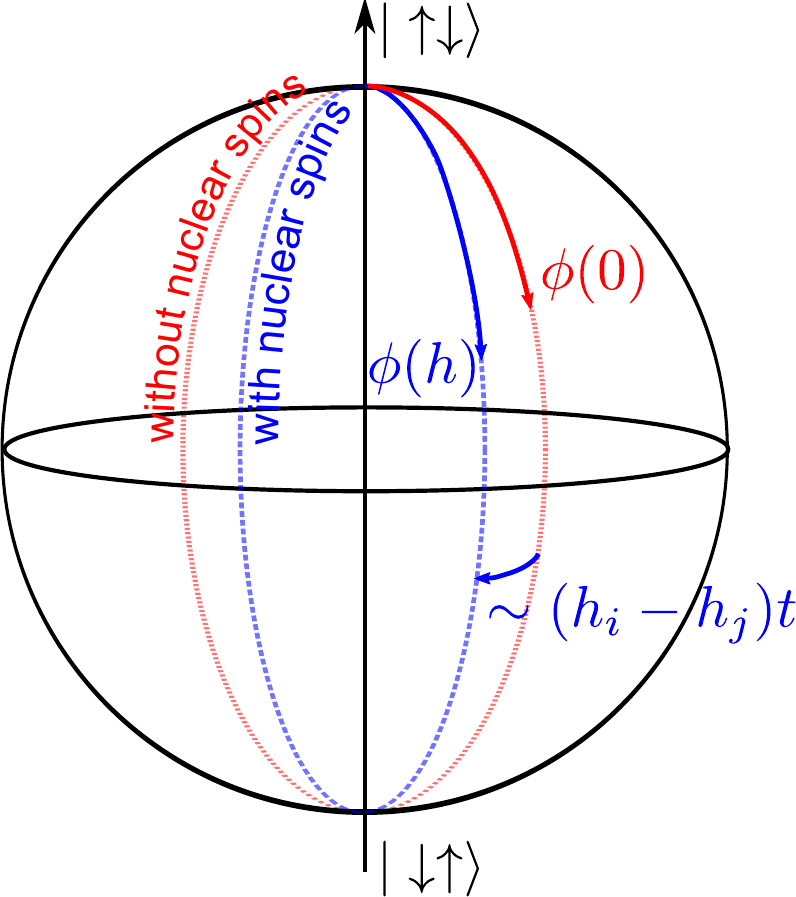}
\caption{Effect of the Overhauser fields $h_i$ and $h_j$ on the time evolution of the two-electron spin state:
change of the interaction phase and precessing of the rotation axis in the x-y plane with $(h_i-h_j)t$}.
\label{SP}
\end{figure}

The unitary time evolution operator of the interaction between two electron spins is
\begin{equation}
 \tilde{U}(\phi) = \exp\left[-i \int \tilde{H}_{\rm int}^{ij} dt\right],
\label{U}
\end{equation}
with $\phi = \int \tilde{g}_{ij}(t)\, dt$.
The time evolution operator in original frame is then
\begin{eqnarray}
%\hspace{-2cm}
U(\phi)&=&R \,\tilde{U}(\phi)R^{\dagger}=\hspace{2.5cm} \\
&=&\begin{pmatrix}
1 & 0& 0& 0\\
0 & \cos (\phi) & i\sin(\phi) {\rm e}^{i(h_i-h_j)t} & 0\\
0 & i\sin(\phi) {\rm e}^{-i(h_i-h_j)t}& \cos(\phi)& 0\\                                  
 0 & 0 & 0 & 1                          
\end{pmatrix}. \nonumber
\end{eqnarray}
In the subspace $\{|\uparrow\downarrow\rangle,\,|\downarrow\uparrow\rangle\} $ the operator $U(\phi)$ acts as 
a rotation $\exp(i\,\phi\,\boldsymbol{\sigma}\cdot\hat{n})$ with $\hat{n}=(-\cos\{(h_i-h_j)t\}, \sin\{(h_i-h_j)t\}, 0)$.
Thus the changed time evolution of the two-electron spin state due to hyperfine interaction can be interpreted as a modified rotation in the mentioned subspace (s. Fig. \ref{SP}). 
There are two distinct effects due to the nuclear spins. First, the interaction phase becomes Overhauser field dependent $\phi=\phi(h_i, h_j)$ and, secondly, the rotation axis starts to precess
in the x-y plane with $(h_i-h_j)t$. The second effect has the maximal contribution when $\phi(0)=m\tfrac\pi2$, where
$\phi(0)=\tilde{g}_{ij}(h_i=0, h_j=0)\, t\equiv \tilde{g}_{ij}(0)\, t$ and where $m$ is an integer. When $\phi(0)=m\pi$, the trajectory of the two-electron spin state affected by nuclear spins coincides with an unaffected one.

The two-qubit CNOT operation can be implemented as a sequence of single spin rotations combined 
with the unitary time evolution operator \cite{imamoglu_prl99}:
\begin{equation}
\begin{split}
 U_{\rm CNOT}=\ue^{-i\tfrac\pi4 \sigma^j_z}\, \ue^{-i \tfrac\pi4}\,
\ue^{-i\tfrac\pi3\hat{n}_i\cdot\boldsymbol{\sigma}_i} \,
\ue^{-i\tfrac\pi3\hat{n}_j\cdot\boldsymbol{\sigma}_j}\times \\
\times U\left(\tfrac\pi4\right)\, \ue^{-i\tfrac\pi2\sigma^i_z} \,  U\left(\tfrac\pi4\right)\,
\ue^{-i\tfrac\pi4\sigma^i_y}\, \ue^{-i\tfrac\pi4\sigma^j_y} \, \ue^{i\tfrac\pi4\sigma^j_z},
\end{split}
\end{equation}
where $\hat{n}_i=(1, 1, -1)/\sqrt{3}$, $\hat{n}_j=(-1, 1, 1)/\sqrt{3}$.

We calculate the fidelity of the generated unitary gate
in the presence of the Overhauser field with respect to the ideal unitary gate 
(without Overhauser field) in order to quantify the effect of the nuclear spins.
This fidelity reflects the difference in final state after the gate operation, averaged
over pure input quantum states \cite{petersen_pra2007}. 
The average 
fidelity for operators acting in 4-dimensional Hilbert space is given by \cite{ghoish_pra2010}
\begin{equation}
 F(\hat{O}(0), \hat{O}(h))=\frac{4+|\mbox{Tr}(\hat{O}^{\dagger}(0)\hat{O}(h))|^2}{20},
\label{generalF}
\end{equation}
where $\hat{O}(h)$ and $\hat{O}(0)$ denote the unitary operator with and without Overhauser field dependence.
Next, the average fidelity Eq.~(\ref{generalF}) must be averaged over the nuclear spin field distributions in 
the quantum dots $i$ and $j$ to extract the Overhauser field dependence of the fidelities for the time 
evolution and the CNOT operator.

\section{Fidelities in second order hyperfine interaction}
Starting from the microscopic model Eq.~(\ref{eh-Hamiltonian}) and 
performing two subsequent  Schrieffer-Wolff transformations
\cite{imamoglu_prl99} 
while including the hyperfine interaction, we find the effective 
spin-spin interaction Hamiltonian as described above. 
Using Eqs.~(\ref{geff}), (\ref{Deltac}), (\ref{DeltaL}),  
 (\ref{gij}), and (\ref{Delta}), we find that the effective spin-spin
coupling depends on the Overhauser fields $h_i$ and $h_j$ in the quantum dots,
\begin{eqnarray}
\label{gfull}
\tilde{g}_{ij}(h_i,h_j)&=&\frac{g^2_{c}\Omega^i_L\Omega^j_L}{16}
\left(\frac{1}{\Delta_i+h_i}+\frac{1}{\Delta_j+h_j}\right)\nonumber\\
&&\times\left(\frac{1}{\Delta_c^i+h_i/2}+
\frac{1}{\Delta^i_L-h_i/2}\right)\\
&&\times\left(\frac{1}{\Delta_c^j+h_j/2}+\frac{1}{\Delta^j_L-h_j/2}\right).\nonumber
\end{eqnarray}
We assume that the fluctuations of the Overhauser field around its zero mean value are small, 
since $h\sim A\sim 10^{-5}$ eV. Therefore we can investigate analytically the fidelities to the second 
order of the Overhauser field. Since the interaction Hamiltonian (\ref{Hint}) is nuclear field 
dependent partly through the coupling coefficient
$\tilde{g}_{ij}$, we need to expand it with respect to the Overhauser field in both coupled quantum dots,
\begin{eqnarray}
\frac{\tilde{g}_{ij}(h_i,h_j)}{\tilde{g}_{ij}(0)}&=&1+\frac{h_ih_j}{4}
\left(\frac{\Delta^2}{\Delta_c^{i}\Delta_L^i\Delta_c^{j}\Delta_L^j}-\frac{1}{\Delta_c^{i}\Delta_L^i}
-\frac{1}{\Delta_c^{j}\Delta_L^j}\right) \nonumber \\
& & + \sum_{l=i,j}\left[\frac{h_l}{2}\left(\frac{\Delta}{\Delta_c^l\Delta_L^l}-
\frac{1}{\Delta}\right)\right.\nonumber\\
& & \left.+\frac{h_l^2}{4}\left(\frac{\Delta^2}{\Delta_c^{l\,2}\Delta_L^{l\,2}}+
\frac{2}{\Delta^2}\right)\right]+\mathcal{O}(h^3) \label{gi}.
\end{eqnarray}
The trace of the product of the perfect time evolution operator
and the one with included hyperfine interaction is 
$\mbox{Tr}[U^{\dagger}(0)\,U(h)]=2\{1+\cos[\phi(0)]\cos(\phi)+\cos[(h_i-h_j)t]\sin[\phi(0)]\sin(\phi)\}$.
We obtain for the average fidelity between those two operators to the second order of the Overhauser
fields $h_i$ and $h_j$, 
\begin{eqnarray}
 F(U)&&=1-\frac{2}{5}\phi^2(0)({\rm D}_i^2 h_i^2+{\rm D}_j^2 h_j^2)\nonumber\\
 &&-\frac{2}{5}\sin^2
 [\phi(0)]\frac{\phi^2(0)}{\tilde{g}_{ij}^2(0)}(h_i-h_j)^2+\mathcal{O}(h^3),
\label{Fa}
\end{eqnarray}
where ${\rm D}_{l}=(\Delta/\Delta_c^l\Delta_L^l-1/\Delta)/2$, $l=i,j$,
are the first-order coefficients in the Overhauser field in 
Eq.~(\ref{gi}), and the time was expressed through the interaction phase $\phi(0)=\tilde{g}_{ij}(0)\, t$. 

\begin{figure}
\includegraphics[width=0.45\textwidth]{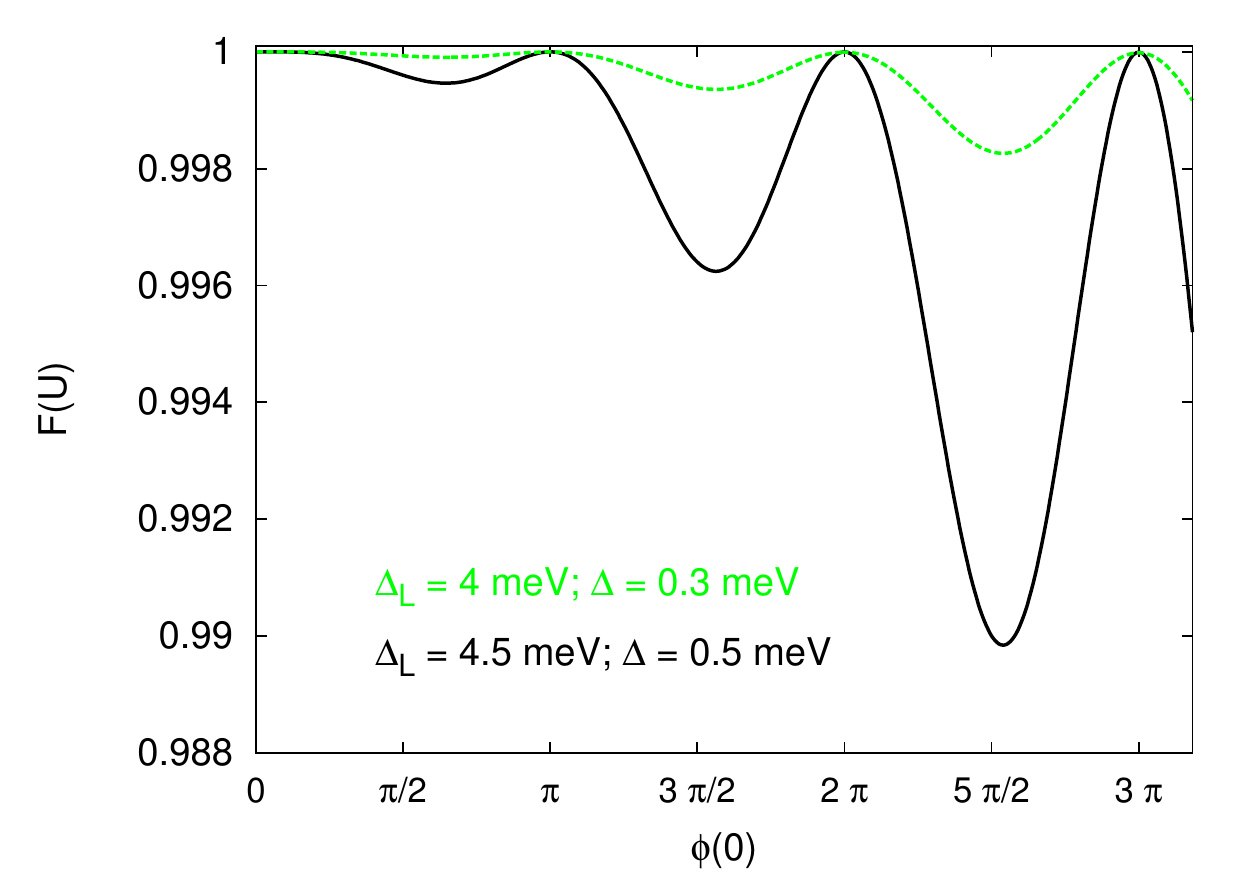}
\caption{Fidelity $F$ of the unitary time evolution operator $U$ generated by the 
XY interaction in the presence of nuclear spins as a function of the interaction phase
$\phi(0)\propto t$, where $t$ denotes the interaction time.
Both lines show the analytical result calculated in second order perturbation
theory in the nuclear field. The fidelity can be increased by tuning the parameters $\Delta$ and $\Delta_L$. For smaller $\Delta$ and $\Delta_L$, the fidelity increases due to faster gate operation. The other parameters are held fixed and are given in the text.  }
\label{FAPlot}
\end{figure}

Averaging of the fidelity (\ref{Fa}) over both Overhauser fields, we find
\begin{eqnarray}
\langle F(U)\rangle_{h_i, h_j} &=&
 1-\frac{2}{5}\phi^2(0)\,\sigma^2({\rm D}_i^2+{\rm D}_j^2) \nonumber\\
 &&-\frac{4}{5}\frac{\phi^2(0)}{\tilde{g}_{ij}^2(0)}\sin^2[\phi(0)]\sigma^2+\mathcal{O}(h^3).
\label{FH}
\end{eqnarray}
Assuming the parameters 
$\Delta_L$ = 4.5 meV, $\Delta_c$ = 5 meV, $g_c$ = 0.5 meV, $\Omega_L^i$ = $\Omega_L^j=$ 1 meV and $\Delta$ = 0.5 meV, 
such that ${\rm D}\equiv {\rm D}_i={\rm D}_j=-0.99\times10^{3}\mbox{ eV}^{-1}$and $\tilde{g}_{ij} = 0.02$ meV, the average 
fidelity given by Eq.~(\ref{FH}) behaves as function of $\phi(0)$ as shown in Fig.~\ref{FAPlot}.
The value of $\langle F(U)\rangle_{h}$ by $\phi(0)=\pi/4$ is equal to $0.99995$. 

\begin{figure}
\includegraphics[width=0.45\textwidth]{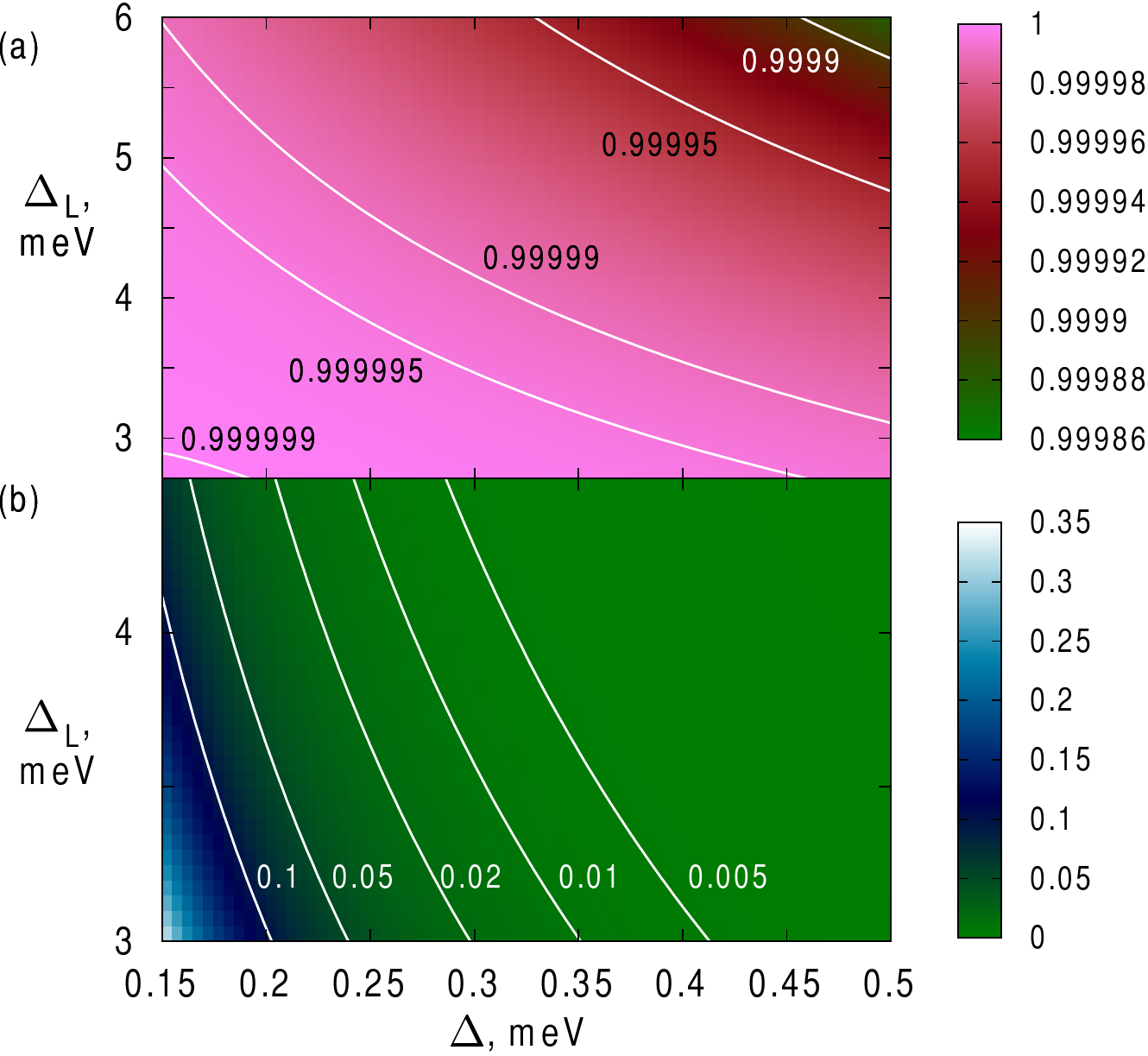}
\caption{a) The average fidelity $F$ of the unitary time evolution operator $U$ for $\phi(0)=\pi/4$ calculated in second order of the 
nuclear field as function of the laser 
detuning $\Delta_L$ and two-photon detuning $\Delta$. 
An adjustment of these parameters can lead to an increase of the fidelity or to its reduction. 
However, the choice of the optimal 
parameters is limited by the conditions of the applied formalism. 
Further details are given in the text
b) The ratio between two contributing mechanism to reduction of the fidelity ${\rm D}^2(\Delta, \Delta_L)\cdot \tilde{g}^2_{ij}(\Delta, \Delta_L)$. It increases for smaller detunings and consequently for improved fidelities.  }
\label{FPar}
\end{figure}

As we can see from the formula (\ref{FH}), the 
two effects induced by nuclear spins decrease the fidelity with corresponding
strengths: the changing of the electron
spin-spin coupling strength and resonance conditions $\propto {\rm D}^2$ 
and the precession of the rotation axis of the two-electron spin state $\propto 1/\tilde{g}_{ij}^2(0)$. 
These are determined in Eqs. (\ref{geff}), (\ref{gfull}), and (\ref{gi}) and depend on 
tunable parameters such as $\Delta$ and $\Delta_L$ (we can express $\Delta_c=\Delta_L+\Delta$). The 
cavity detuning can be tuned accordingly, e.\,g. in photonic two-dimentional slab microcavities using Xe condensation \cite{khitrova}.   
The contributions to fidelity reduction from terms $\propto {\rm D}^2$ and $\propto 1/\tilde{g}_{ij}^2(0)$ are not of the same order (s. Fig. \ref{FPar} b), 
${\rm D}^2\le  0.1/\tilde{g}^2_{ij}(0)$. But we still keep the terms $\propto {\rm D}^2$ in our analytical calculations, because they give the upper boundary 
of the fidelity, when $\sin[\phi(0)]=0$. By changing the laser detuning $\Delta_L$ and the detuning $\Delta$ we can 
adjust the values ${\rm D}$ and $\tilde{g}_{ij}$ and thus tune the fidelity itself.

The dependence of the fidelity (for $\phi(0)=\tfrac\pi4$) in respect to $\Delta$ and $\Delta_L$ is shown in Fig. \ref{FPar}(a). 
It indicates that fidelity can be improved considerably by changing the detunings.
However by choosing better parameters,
some assumptions of the applied theory should be conserved, so that $\tilde{g}_{ij}\ll \Delta$ and $\Delta_L (\Delta_c)\gg g_c$.
One can tune the intensity of applied control laser and rotate the electron spins faster, so that they interact faster as well 
and ``feel" the dephasing in a lesser extent. However the coupling strength $\tilde{g}_{ij}(t)$ depends quadratically on laser
intensity and just slight enhancement of Rabi frequency $\Omega_L$ will make it of the same order as $\Delta$.
Therefore, to improve the fidelity and to keep these relations true one can use e. g. $\Delta$ = 0.3 meV and $\Delta_L$ = 4 meV.
With this choise, the average fidelity for the gate $U(\frac\pi4)$ is increased to 0.999991. The fidelity decays considerably slower in this case as well (Fig. \ref{FAPlot}).
The formula Eq.~(\ref{generalF}) can be applied for calculating the average fidelity of the CNOT operation.
If we neglect the imperfections of the single qubit operations in the CNOT operator sequence
and focus only on the decoherence due to the interaction part, then
we find for the trace distance between a perfect CNOT operator and CNOT 
operator with nuclear spins,
\begin{eqnarray}
 \mbox{Tr}&&[U^{\dagger}_{\rm CNOT}\,U_{\rm CNOT}(h)]=\\
 &&2\left(1+\cos\{2[\phi-\phi(0)]\}\cos\Bigg[(h_i-h_j)\frac{\phi(0)}{\tilde{g}_{ij}(0)}\Bigg]\right)\big|_{\phi(0)=\tfrac\pi4}.\nonumber
\end{eqnarray}
The CNOT average fidelity in second order in the Overhauser field is then
\begin{eqnarray}
 F^{\rm CNOT}&=&1-\frac{16}{5} \Big[ \phi(0)=\tfrac\pi4\Big]^2{\rm D}^2\sigma^2 \nonumber\\
 &&-\frac{4}{5}\Big[\phi(0)=\tfrac\pi4\Big]^2\frac{\sigma^2}{\tilde{g}^2_{ij}(0)}+ \mathcal{O}(h^3).
\label{FCNOT}
\end{eqnarray}
We find that $F^{\rm CNOT} = 0.9999$ by using $\Delta$ = 0.5 meV and $\Delta_L$ = 4.5 meV, and 
$F^{\rm CNOT} = 0.999983$ with $\Delta$ = 0.3 meV and $\Delta_L$ = 4 meV, 
which is smaller than the fidelity of 
a single $U(\tfrac\pi4)$ operation. 

\begin{figure}
\includegraphics[width=0.48\textwidth]{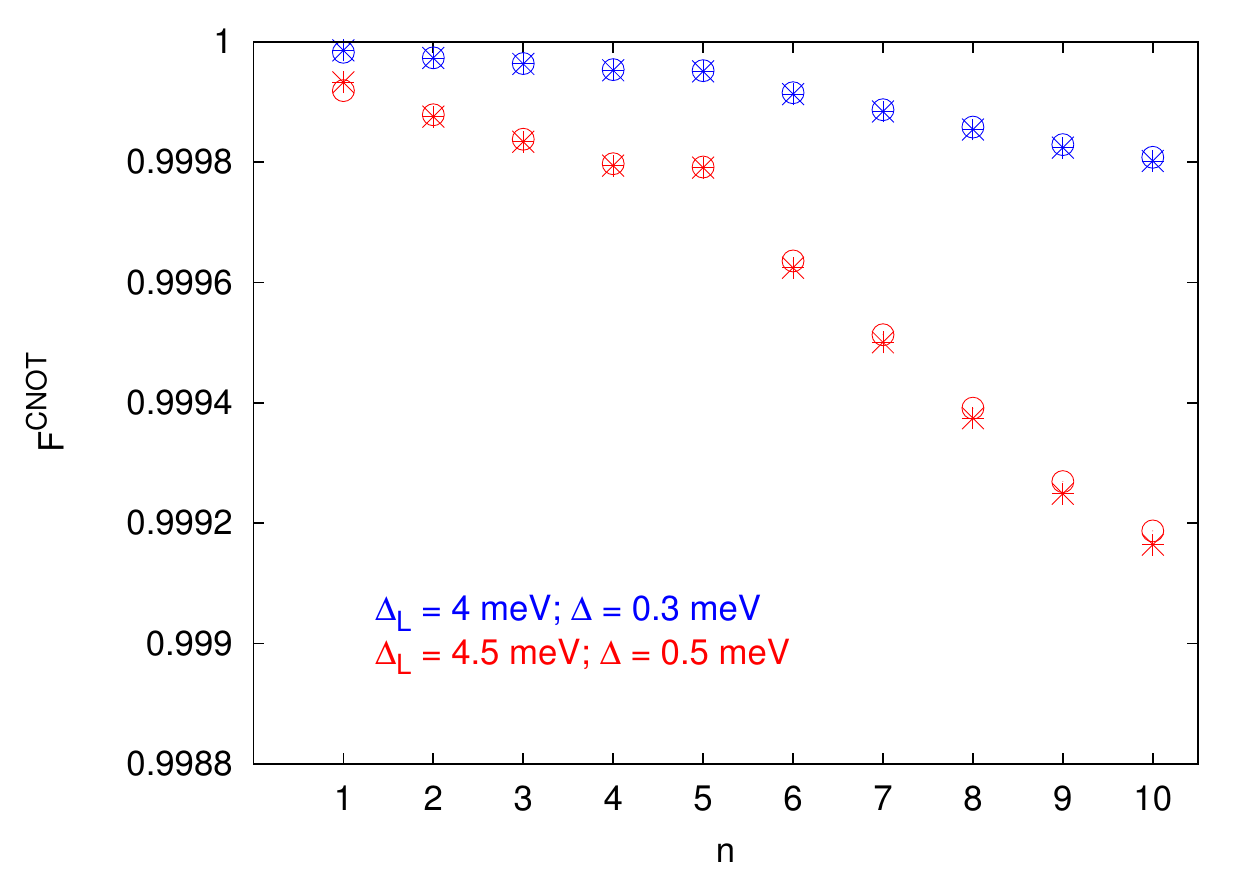}
\caption{The analytical and numerical results for fidelity of $n$ subsequent CNOT operations in the presence of 
nuclear spins for different detunings. The analytically calculated values are represented by circles and the numerically
calculated ones by stars.}
\label{FCPlot}
\end{figure}

For the  average fidelity for further applied CNOT operations, we find
\begin{equation}
 F^{\rm CNOT}(n)=\frac{4+|\mbox{Tr}\{(U^{\dagger}_{\rm CNOT})^n\,[U_{CNOT}(h)]^n\}|^2}{20}.
\label{FN}
\end{equation}
The fidelity $F^{\rm CNOT}(n)$  averaged over the nuclear field (Fig.~\ref{FCPlot})
is reduced  after each application and can be increased by a better choice of the detuning parameters. 
Thus for the last set of parameters only after seven CNOTs, the fidelity drops below typical error correction thresholds ($<$ 0.9999)  \cite{nielsen&chuang}.
\section{Numerically exact fidelities}
For large values of the parameter $\phi(0)$ we evaluate the averaging of the fidelities over nuclear fields 
without expanding $|\mbox{Tr}(U(0)^{\dagger}U(h))|^2$.
\begin{figure}
\includegraphics[width=0.45\textwidth]{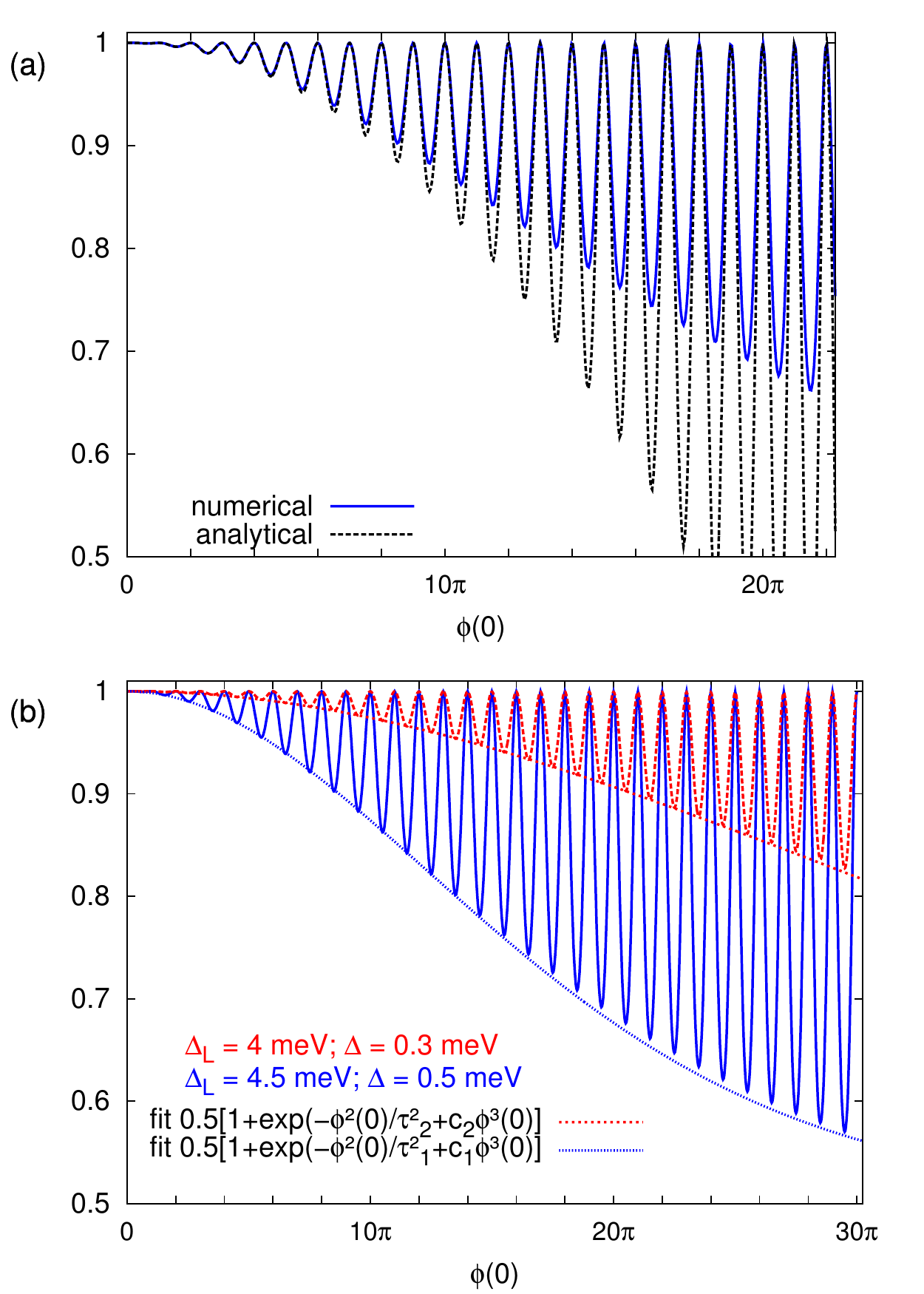}
\caption{Numerically calculated average fidelity of the time evolution
operator for large $\phi(0)$. (a) The comparison between analytical and numerical calculations for  $\Delta_L$ = 4.5 meV
and $\Delta$ = 0.3 meV. The analytically obtained fidelity diverges for large $\phi(0)$ while the numerical fidelity saturates to 0.5.
(b) Comparison between two numerically calculated fidelities for the gate $U(\phi)$ with different detunings.}
\label{FullPlot}
\end{figure}
The results of the numerical averaging are shown in the Fig.~\ref{FullPlot}. 
The comparison to the analytical result in Fig.~\ref{FullPlot}(a) reveals that our
analytical model can predict the behavior of the fidelity only up to a certain value
of $\phi(0)$. From Fig.~\ref{FullPlot}(a) one can see that around $\phi(0)\simeq$ 12 the 
numerical and analytical data begin to differ. For large $\phi(0)$, the analytical data even
diverges into negative values. The numerical data on the other hand converges to 1/2.
The numerical evaluation of the fidelity for different sets of detunings [Fig.~\ref{FullPlot} (b)] confirms 
that the tuning of the transition parameters can improve the fidelity significantly. The numerical curve oscillates with
the same period $2\pi$ as the analytical one. The maxima correspond to the situation, when
the fidelity is decreased only due to the modified interaction phase between the electron spin and there is no effect 
of the changed rotation axes of the two-electron spin state. The maxima correspond to the decay of the fidelity, where 
both Overhauser field induced effects are contributing and the effect of the precession of the rotation axis is maximal.                 

The fidelity curve suggests that the behavior of its lower boundary
could be described by the Gaussian  $\sim\ue^{-\phi(0)^2}$. 
The fidelity envelope to the second order of $\phi(0)$ is obtained from Eq.~(\ref{FH}) by
setting $\sin (\phi(0))^2$ = 1, 
We fit this envelope to
the decay function $\tfrac12[1+\exp(-\phi(0)^2/\tau^2)]$, where $\tau$ is given
through parameters $\sigma$, ${\rm D}$ and $\tilde{g}_{ij}$ from Eq.~(\ref{FH})
and is defined as 
\begin{equation}
\tau=\sqrt{\frac{5}{8}}\frac{\tilde{g}_{ij}(0)}{\sigma\sqrt{1+{\rm D}^2\,\tilde{g}^2_{ij}(0)}}
=\tau(\Delta, \Delta_L).
\end{equation}
The Gaussian function gives a good fit to the minimum points of the fidelity for small $\phi(0)$, but
decays faster for larger $\phi(0)$. The tail of the fidelity can be fitted by 
adding a third-order term to the 
argument of the exponential function,
 $\tfrac12[1+\exp(-\phi(0)^2/\tau^2+c\,\phi(0)^3)]$ with $c$ as a fitting parameter. 
The fits are shown in Fig.~(\ref{FullPlot}) and fit with $\tau_1$ and $c_1$ corresponds 
to the curve with parameters $\Delta_L$ = 4.5 meV and with $\tau_2$ and $c_2$ to numerical curve
with another set of parameters.
The fits yield $c_1 = 9.6\times10^{-7}$ and $c_2 = 5\times10^{-8}$. 

The time dependance of the fidelity can be found by changing from 
interaction phase $\phi(0)$ to time by assuming $\phi(0)=\tilde{g}_{ij}(0)\,t$.
Then for short time dependance of the lower boundary of the fidelity is described by
the function $0.5[1+\exp(-t^2/T^2)]$, where 
\begin{equation}
T= \sqrt{\frac{5}{8}}\frac{\hbar}{\sigma\sqrt{1+{\rm D}^2\,\tilde{g}_{ij}^2(0)}}. 
\end{equation}
Since from Fig.~(\ref{FPar}) we know that ${\rm D}^2\,\tilde{g}_{ij}^2(0) \ll 1$, we can neglect it here. It follows that
$T=\sqrt{\frac{5}{8}}\frac{\hbar}{\sigma}=\sqrt{\frac{5}{8}}\frac{\hbar\sqrt{N}}{A}$, where $N$ is the number of the nuclear spins
and $A$ is hyperfine interaction constant.
This implies that the decay of the fidelity does not dependend on the interaction 
strength of the electron spins and is only affected by the nuclear field distributions. By tuning the transition parameters
we increase the interaction strength between the coupled electron spins to some reasonable value. Thus the spins acquire 
a larger interaction phase within the same time. But the decay time of the time evolution operator fidelity is the same and for 
GaAs ($N = 10^5$, $A$ = 90 $\mu$eV) $T=$ 2 ns. The $T_2$ decay time of a electron spin coupled to the nuclear bath is given by
$T_2=\frac{2\hbar\sqrt{N}}{A}$ \cite{coish_prb2004} and in GaAs is $\approx$ 5 ns. Since both interacting spins are affected 
by the decoherence, the decay of the common interaction phase is faster than of a single one ($T\approx 0.4 \,T_2$).  

The numerically exact CNOT fidelities are presented in Fig. \ref{FCPlot}. The second order approximation in 
Overhauser field gives a good agreement to the numerical results. Since the CNOT operator is defined for a fixed interaction phase,
there is no oscillation behavior and the fidelity decreases for each further applied CNOT gates.   
\section{Conclusions}
We calculated average fidelities for unitary time evolution operator and CNOT operation of two quantum 
dots interacting through an off-resonant cavity mode. We obtained results to second order of 
hyperfine interaction
analytically and to an arbitrary order by averaging the fidelities over the Overhauser field distributions
numerically. Both approaches are in good agreement for small interaction time durations. 
If only the longitudinal component of the Overhauser field in two interacting quantum dots is considered for the hyperfine interaction, then the time evolution of the two-electron spin state changes in two ways (Fig. \ref{SP}),
and both effects contribute differently to the reduction of the fidelity. A prolongation of the decoherence time
of the two-electron state is possible by narrowing the distribution of the Overhauser field \cite{stepanenko_prl2006} [Eq. (\ref{FH}), (\ref{FCNOT})]. The precession of the rotion axis of the two-electron state on the Bloch sphere can be minimized by preparing the the nuclear spin states in each interacting quantum dot in the same certain state and 
minimizing the fluctuations of the Overhauser fields \cite{ribeiro_prl2009}. Since we neglected the hyperfine spin-flip terms in our 
calculations, the obtained results are only valid for the Zeeman fields $\propto$ 1 T.

\section{Acknowledgments}
We acknowledge funding from the DFG within SFB 767.

\end{document}